\documentclass[twocolumn,preprintnumbers,aps]{revtex4}

\usepackage{epsfig}
\usepackage{amsmath,amssymb,amsfonts}

\newcommand{\Sf}{{\cal {S}}}
\newcommand{\be}{\begin{equation}}
\newcommand{\ee}{\end{equation}}
\newcommand{\bea}{\begin{eqnarray}}
\newcommand{\eea}{\end{eqnarray}}

\newcommand{\vk}{{\mathbf k}}
\newcommand{\vx}{{\mathbf x}}

\newcommand{\bq}{\begin{eqnarray}}
\newcommand{\eq}{\end{eqnarray}}
\newcommand{\lp}{\ell_{\mathrm P}}
\def\f{\frac}
\def\g{\gamma}

\def\p{\partial}

\def\heff{H_{\mathrm{eff}}}

\def\ba{\begin{eqnarray}}
\def\ea{\end{eqnarray}}
\def\Hu{\frac{\dot a}{a}}
\def\Hu2{\frac{\dot a^2}{a^2}}

\def\hkl{h_k^{(\lambda)}}
\def\hkli{h_k^{(\lambda)-1}}

\begin{document}
\preprint{IGPG-07/3-3}
\title{Thermal fluctuations in loop cosmology}
\author{Jo\~{a}o Magueijo$^{1,2,3}$, Parampreet Singh$^{4,1}$}
\affiliation{ $^1$Perimeter Institute for Theoretical Physics, 31
Caroline St N, Waterloo N2L 2Y5, Canada\\
$^2$ Canadian Institute for Theoretical Astrophysics,
60 St George St, Toronto M5S 3H8, Canada\\
$^3$ Theoretical Physics Group, Imperial College, Prince Consort
Road, London SW7 2BZ, England\\
$^4$ Institute for Gravitational Physics and Geometry,
The Pennsylvania State University, University Park, PA 16802, USA}

\begin{abstract}
Quantum gravitational effects in loop quantum cosmology lead to a
resolution of the initial singularity and have the potential to solve the
horizon problem and generate a quasi scale-invariant spectrum of
density fluctuations. We consider loop modifications to the behavior
of the inverse scale factor below a critical scale in closed models
and assume a purely thermal origin for the fluctuations.  We show that
the no-go results for scale invariance in classical thermal
models can be evaded even if we just consider modifications to the
background (zeroth order) gravitational dynamics. Since a complete
and systematic treatment of the perturbed Einstein equations
in loop cosmology is
still lacking, we simply parameterize their expected
modifications. These change quantitatively, but not qualitatively, our
conclusions.  We thus urge the community to more fully work out this complex
aspect of loop cosmology, since the full picture
would not only fix the free parameters of the theory,
but also provide a model for a non-inflationary, thermal origin for the
structures of the Universe.
\end{abstract}

\maketitle

\section{Introduction}

The possibility that primordial thermal fluctuations might
seed the structure
of our Universe is an intriguing alternative to
quantum fluctuations in a deSitter phase~\cite{peeb,steph,rob,mag-pog}.
Unfortunately a number of  obstacles present themselves to such an
enterprise. Firstly any thermal scenario should
necessarily be based on a solution of the horizon problem.
This is so that the assumption of thermalization itself makes sense: modes
must start off in causal contact to thermalize and then leave the horizon.
(This is in fact true of any scenario where the fluctuations are
``passive''~\cite{andy,passive}.) A number of solutions to the horizon problem
have been proposed~\cite{infl,hag,vsl} and in this paper we use in effect
a combination of two.

But more importantly thermal scenarios
run against an apparent wall:
the well know fact that thermal fluctuations have a
white-noise spectrum, i.e. spectral index $n_S=0$, rather than the observed
near-scale-invariance, $n_S\approx 1$.
Thus any scenario where the primordial fluctuations result from
a ``snap-shot'' of a thermal bath at a fixed
temperature is doomed. This discouraging result, however, may be
circumvented by noting that the white-noise nature of thermal fluctuations
follows from the extensive nature of the energy.
Reasonable and general
as this assumption might be, it could be violated in the early
Universe, during a phase
ruled by new physics at the Planck or string energy scale.
This has been suggested by at least
two lines of research. In one a gas of strings at the Hagedorn phase is
employed~\cite{hag}. Another invokes a holographic phase in loop
quantum gravity~\cite{holo}.

Yet another solution is to ensure that different modes leave the
horizon and freeze-out at different temperatures. Then,
the equal-temperature spectrum of thermal fluctuations is still
white-noise, but the spectrum of frozen-in fluctuations
imprinted outside the horizon isn't. The actual form of that
spectrum depends on the balance between the size of the mode
leaving the horizon at a given time and the temperature (and thus
the mode's amplitude) at the time the mode is picked out
of the thermal bath, leaves the horizon and freezes-in.
There is some controversy over whether this mechanism may lead
to a scale-invariant spectrum and one
of the purposes of this paper is to clarify the matter.

In Section~\ref{nogo} we provide a model calculation based on a
minimally modified thermal scenario, in which thermal matter is
allowed to have a different equation of state with
$w=p/\rho<-1/3$, but where nothing else is changed. We show that
{\it unless new physics comes into
play, modifying the Einstein equations, the thermodynamical
relations, or some other standard assumption,
in all such scenarios the spectral index is $n_S=4$}. This is
true regardless of $w$, the only free parameter of the model.
This section is supplemented by Appendix~\ref{measures} which
provides all relevant definitions of measures of structure.

Of course it is  natural that new physics does come into play
in the early Universe, and  the
rest of this paper is focused on the potential of loop quantum
cosmology (LQC) to reverse this negative result (for an up to date introductory
 review see \cite{ab-rev} and for early developments in the
field see \cite{mblr}). Modifications to
the the Einstein equations in LQC originate from two sources: the
field strength (curvature) of the Ashtekar connection which is
expressed in terms of holonomies and the
inverse powers of volume in the constraint. These are quantified
by two parameters in the theory. Firstly, the $j$ parameter, which
appears due to tracing over holonomies of the SU(2) connection in both
gravitational and matter parts of the constraint. This parameter
also determines the scale at which modifications to inverse volume
become significant.
Secondly, the $l$ parameter, which arises due to the inverse volume term in
the matter part and quantifies the functional form of this
modification.

So far a complete and consistent quantization of LQC
with the knowledge of physical Hilbert space has only been
performed for $j=1/2$. It shows a physical resolution of the
singularity and leads to the correct classical limit
\cite{aps,aps2,polish-col,apsv,kv-k-1}. Nevertheless, these
investigations have provided valuable insights on existence of
effective
Hamiltonian which is an excellent approximation to the underlying
quantum dynamics and which can be generalized to higher $j$ in
particular for the regime where the Hubble rate is small compared
to Planck scale.
In this regime, modifications to the Einstein equations are
primarily due to corrections to the inverse volume operator.
Though these modifications unfortunately cannot be tied to a
curvature/energy density scale in non-compact flat models
\cite{aps2,kv-k-1}, they are related to the scale of intrinsic
curvature in closed models and as we will show also lead to a bounce of the closed
universe at a critical scale factor \cite{fn1}.
Our analysis will be carried out in this
framework.

In Sec ~\ref{lqc} we present this effective Hamiltonian and the
derivation of the background dynamical equations. To keep the
paper suitable for a phenomenology oriented readership, all
the necessary details of loop quantization are relegated to
Appendix~\ref{lqc-app}, where we give particular emphasis to the
origin of the $j$ and $l$ parameters with resulting correction terms
in the new improved quantization of LQC \cite{aps2,polish-col,apsv,kv-k-1} and the role of different
modifications to the Einstein's equations.
Equipped with the background dynamics in LQC, in Section \ref{thermal}  we
then investigate the spectrum of thermal fluctuations.
We demonstrate that the modified dynamics in
LQC evades the no-go result of Sec.~\ref{nogo}.
The constraint on the near scale
invariance of fluctuations translates into the requirement on the
effective equation of state in loop cosmology, $w \approx -2/3$.
This can be obtained by a suitable choice of the $l$ parameter.

Since work on inclusion of inhomogenities in loop cosmology is at a very early
stage of development~\cite{struct} (and various technical aspects,
in particular those relevant for the regime of
interest, are yet to be understood) we adopt a phenomenological
approach and parameterize the expected modifications to the perturbed Einstein
equations.  We show that expected corrections in fact make the above
result stronger.

More work needs to be carried out by the community
until our calculations may be converted into specific constraints.
But, as we summarize in
Section~\ref{concs}, in this paper we are able
to provide a list of what exactly needs to be worked out
so that $n_S\approx 1$ is converted into a constraint upon the
free parameters of LQC.

\section{The base calculation and the No-Go result
in classical physics}\label{nogo}
Let us consider a thermal scenario in which the only effect of new
physics is to change the equation of state of thermal matter.
This certainly happens in theories with deformed dispersion
relations~\cite{steph,rob}, and also in LQC~\cite{ps-disp}.
We then assume that
statistical physics, the gravitational equations and  a few basic
thermodynamical relations are not modified. This won't
necessarily happen
in loop cosmology, but the way in which the modifications
arise is not yet fully understood. It is therefore interesting
to provide a base calculation, assuming no changes, as a blueprint
for further work.

The calculation follows three steps.

\subsection{The fixed temperature power spectrum}
The first step is to compute the fixed temperature power spectrum.
This turns out to depend only on the specific heat at constant volume
(a result that has been known since the XIX century, and is now
textbook material~\cite{kittel}). The spectrum is generally
white noise, a fact that can be directly traced
to the extensive nature of the energy, i.e. to the fact that the energy
inside a given region is proportional to its volume.

The derivation is very general. Consider the partition function
\begin{equation}
Z={\sum_r} e^{-\beta E_r} \ ,
\end{equation}
where $\beta = T^{-1}$. The total (matter) energy $U$ inside a volume $V$ is
given by:
\begin{equation}
U={\langle E\rangle}={{\sum _r} E_r e^{-\beta E_r} \over {\sum_r}
e^{-\beta E_r}}=-{d\log Z\over d\beta}
\end{equation}
and its variance by
\begin{equation}\label{varE}
\sigma^2_E={\langle E^2\rangle}-{\langle E\rangle}^2={d^2\log
Z\over d\beta^2}= -{dU \over d\beta}=T^2c_V
\end{equation}
where $c_V$ is the specific heat at constant volume.
If the energy is extensive  then $U=\rho V$,
with energy density $\rho=\rho(T)$, that is $U$ is
proportional to the volume.
The spectral index and amplitude may now be found by means of the
following tools (for the closed model we will be interested in the
regime where intrinsic curvature terms can be neglected):
\begin{itemize}
\item  The Poisson equation
\be\label{poisson1}
k^2\Phi=4\pi G a^2 \rho \delta= 4\pi G a^2 \delta\rho,
\ee
relating the gravitational potential $\Phi$
and the density fluctuations. Outside the horizon there may be gauge issues,
but this relation certainly holds for subhorizon modes.
\item
The proportionality between the variance $\sigma^2_g(R)$ in a quantity
$g$ defined in position space and smeared on a scale $R$, and the
``dimensionless power spectrum'':
\be\label{sigPapprox1}
\sigma^2_g(R)={\langle \delta g^2\rangle}_R\approx {\cal
P}_g(k_R=a/R) \ee
(see Appendix~\ref{measures} for definitions; note that here $k$
is comoving, but $R$ is a proper size).
The spectral index $n_S$ is defined from ${\cal P}_\Phi=A^2k^{n_S-1}$.
Formula (\ref{sigPapprox1}) has not been questioned for $n_S<1$;
but see~\cite{joy,dav}.
\end{itemize}
Then using ({\ref{varE}) we have
\be
{\langle \delta\rho^2\rangle}_R={1\over R^6}{\langle \delta E^2\rangle}_R
={\sigma^2_E(R)\over R^6}={T^2\over R^6}c_V\, .
\ee
Combining this result with
(\ref{poisson1}) and (\ref{sigPapprox1}) (for $g=\delta\rho$)
we thus conclude
\be
{\cal P}_\Phi\sim {a^4{\cal P}_{\delta\rho}\over k^4}\sim {a^4\over k^4}
{\left[{T^2\over R^6} c_V\right]}_{R=\frac{a}{k}}=
{k^2\over a^2}T^2 [c_V]_{R=\frac{a}{k}} ~.
\ee
Using the extensive nature of the energy we have $c_V=\rho'(T)R^3$,
so finally
\be
{\cal P}_\Phi\sim {a\over k}T^2 \rho'\, .
\ee
The fluctuations are therefore white noise ($n_S=0$), and have an amplitude
that only depends on the Stephan-Boltzmann law, relating energy density
and temperature. We shall assume a Stephan-Boltzmann law
of the type $\rho\propto T^\zeta$, where $\zeta$ is a parameter
subject to thermodynamical constraints to be discussed later.

\subsection{Frozen-in power spectrum}
We assume that comoving scales $k$ start thermalized and inside the horizon,
and then leave the horizon, with (first) crossing defined by
$k=aH$. This requires either accelerated expansion~\cite{infl}, a
loitering phase~\cite{hag}, a decreasing speed of light~\cite{vsl},  a
bouncing scenario, or a combination thereof.
We use the first mechanism, so that
the equation of state satisfies $w<-1/3$.

As the $\Phi$ modes leave the horizon their amplitude gets fixed at whatever
thermal amplitude they have at crossing, that is:
\be\label{xing}
{\cal P}_\Phi(k)\sim {\left[{a\over k}T^2 \rho'\right]}_{k=aH}\, .
\ee
Since different modes freeze at
different temperatures the spectrum left outside the horizon won't
be white noise.

Using the Friedman equation $H^2\propto\rho$ we can rewrite (\ref{xing})
as
\be\label{PT}
{\cal P}_\Phi(k)\sim {\left[T^2\rho'\over \sqrt{\rho}\right]}_{k=aH}
\ee
where $k=aH$ specifies a relation between a given comoving $k$ leaving
the horizon at a given time, and the temperature, thereby allowing the
inversion of the right hand side as a function of $k$.
Eqn.~(\ref{PT}) implies:
\be\label{nPofT}
{d\ln {\cal P}_\Phi\over d\ln T}=1+{\zeta\over 2} ~.
\ee
The relation between $k=aH$ and the temperature, however,
depends on both the equation of state $p=w\rho$ and
$\rho\propto T^\zeta$. Using the Friedmann equation we have
$k=aH\propto a\sqrt{\rho}$, and  since $\rho\propto 1/a^{3(1+w)}$,
we may derive
\be\label{aofT}
a\propto T^{-\zeta\over 3(1+w)}.
\ee
Therefore:
\be\label{nkofT}
{d\ln k\over d\ln T}= {-\zeta\over 3(1+w)}+{\zeta\over 2}
={\zeta(1+3w)\over 6(1+w)} ~.
\ee
We can now compute the spectral index as
\be\label{n-1}
n_S-1={d\ln {\cal P}_\Phi\over d\ln k}=
{d\ln {\cal P}_\Phi\over d\ln T}{d\ln T\over d\ln k}
=3{2+\zeta\over \zeta}{1+w\over 1+3w}
\ee
but note that the condition $w<-1/3$ (or that $k=aH$ increases in time)
is necessary for this formula to make sense.

Two promising regions of parameter space stand out. Firstly
$\zeta=-2$, that is $\rho\propto1/T^2$; this may lead to
scale-invariance because the amplitude of the frozen-in
thermal fluctuations does not depend on the temperature in this case
(c.f. Eqn.\ref{PT}). Secondly $w=-1$; one can see that this
could lead to scale-invariance because $\rho$ does not change
(it behaves like a cosmological constant), and so neither
does the temperature or amplitude of the fluctuations as they leave
the horizon.

However further conditions apply. Regarding the  first case we
have to check that
$w<-1/3$ is possible, so that modes do leave the horizon.
With respect to the latter, we should additionally
have $\zeta\neq 0$ (or $\zeta\neq\infty$), so that there are fluctuations
at all (and they are not infinite).
Unfortunately closer inspection shows that these conditions
cannot be met.

\subsection{Thermodynamical constraints}
It's been noted~\cite{verl,youm,dias} that the equation of state $p=w\rho$
and the Stephan-Boltzmann law $\rho=\rho(T)$ are linked by a
thermodynamical relation. The argument assumes that energy and entropy
are extensive. Consider the first law of
thermodynamics:
\be\label{1stlaw}
dU=-PdV+TdS ~.
\ee
If the energy $U$ and entropy $S$ are extensive, then
$U(\lambda V,\lambda S)=\lambda U(V,S)$. Taking a derivative with
respect to $\lambda$ at $\lambda=1$, and using(\ref{1stlaw})
we arrive at the Euler relation
\be\label{euler}
U=-PV+TS
\ee
so that defining $\rho=U/V$ and entropy density $s=S/V$ we have
\be
s={P+\rho\over T} ~.
\ee
We can now prove that $s=dP/dT$ in a variety of ways, e.g. introducing
the free energy $F=U-TS=F(V,T)$, so that $dF=-PdV-SdT$. This leads
to the integrability condition:
\be
s={\left(\partial S\over\partial V\right)}_T={\left(\partial
P\over
\partial T \right)}_V ~.
\ee
Thus the expression
\be\label{prhot}
{dP\over dT}={P+\rho\over T} ~.
\ee
If $w$ is a constant we obtain that $\rho\propto T^\zeta$
with
\be\label{zetaw}
\zeta=1+{1\over w} ~.
\ee
The trouble is that this relation implies that $\zeta=0$ for $w=-1$:
``deformed'' radiation may behave like a cosmological constant,
but then the specific heat vanishes and there are no thermal
fluctuations at all. This is an interesting
result but kills the second candidate for scale-invariance
proposed above.

The first candidate is killed by noting that $\zeta=-2$ implies $w=-1/3$,
that is a Milne Universe. This is merely
a borderline case for solving to the horizon problem: the comoving
horizon does not increase but neither does it increase.

One might expect that models near these two can bypass these problems
and display if not strict scale-invariance, at least approximate
scale-invariance. However this is not the case. Inserting (\ref{zetaw})
into (\ref{n-1}) a simple algebraic calculation shows that
$w$ (or $\zeta$) cancel out and we are left with
\be
n_S=4
\ee
for {\it all} model parameters.

Therefore one needs further new physics to bypass this negative result.
Presumably the double branched dispersion relations considered
in~\cite{rob} are behind the fact that $w=-1$, $\zeta=1$ is
possible (in contradiction with (\ref{zetaw}). We now examine the
way modifications to the dynamics in LQC reverse these results.

\section{Effective Dynamics in loop quantum cosmology}\label{lqc}

The phase space in LQC
consists of the geometrical variables -- the connection $c$ and
the triad $p$ -- and the matter variables, which for a scalar field
will be $\phi$ and its canonical momenta $p_\phi$. The triad is
related to the scale factor as $p = a^2 =
V^{2/3}$. On the classical solutions of GR for closed model, $c$ is related to the
time derivative of scale factor as $c = \gamma \dot a + 1$ where
$\gamma \approx 0.2375$ is the Barbero-Immirzi parameter. The
connection and triad are canonically conjugate satisfying
\be\label{cpbp}
\{c,p\} \, = \, \f{8 \pi G \gamma}{3} ~.
\ee
For the closed model, in the regime where the
Hubble rate is small compared to
Planck scale we can write an effective Hamiltonian which encodes
the modifications to the inverse scale factor below a critical
scale $a_\star$ (parameterized by $j$ (\ref{astar})) in terms of functions ${\cal S}$ (Eq.(\ref{slta*})) and
$D_l$ (Eq.(\ref{dlta*})):

\be\label{effhamk1}
\heff = -\f{3}{8\pi G \g^2} \, \Sf a \, \left((c - 1)^2 + \gamma^2
\right) +
H_m ~
\ee
where $H_m$ is the matter Hamiltonian obtained after inverse
volume modifications using (\ref{sjdj}). For a massive scalar field it
is
\be
H_m = \f{1}{2} \, D_l \, \f{p_\phi^2}{a^3} \, + \, a^3 \,
{\cal V}(\phi)~.
\ee

Dynamics can now be obtained by the use of Hamilton's equations.
In order to obtain the modified Friedman equation we first
evaluate
\bq \label{pdot}
\dot p &=& \nonumber  \{p,\heff\} = - \f{8 \pi G \g}{3} \, \f{\p
\heff}{\p c} \\
&=& \f{2 \Sf a}{\g} \, (c - 1)
\eq
and then using it in $\heff \approx 0$:
\bq\label{mod-fried}
\f{\dot a^2}{a^2} &=& \nonumber \f{8 \pi G}{3} \, \Sf\,
\f{H_m}{a^3} - \f{\Sf^2}{a^2} \\
&=& \f{8 \pi G}{3} \, \Sf \rho_{sc} - \f{\Sf^2}{a^2}
\eq
where $\rho_{sc} = H_m/a^3$ denotes the modified energy density
\cite{fn2} (we follow the conventions of Ref. \cite{ps-disp}). It can be easily seen that a bounce
occurs for $a < a_\star$ when $(8 \pi G/3) \rho_{sc} = \Sf/a^2$ which is possible
due to form of ${\cal S}$ and $D_l$ in this regime. Also, for $a
\gg a_\star$, ${\cal S}$ and $D_l$ approach unity yielding us the
classical Friedman dynamics.

We can obtain the modified Raychaudhuri equation using
Hamilton's equation for $c$:
\be
\dot c = \{c, H_{\mathrm{eff}} \} = \f{8 \pi G \g}{3} \,\f{\p \heff}{\p p}
\ee
and the expression for the modified pressure
\be
P_{sc}  = -\f{\p H_M}{\p V} = - \f{2}{3} p^{-1/2} \, \f{\p H_M}{\p p}
~.
\ee
These equations lead to
\be
\dot c = -\f{1}{2 \g} \, \left((c - 1)^2 + \g^2\right) \,
\left(\f{\dot \Sf}{\dot a} + \f{\Sf}{a} \right) - 4 \pi G \g \, a
P_{sc} ~
\ee
which combined with the time derivative of Eq.(\ref{pdot}) result in 
the modified Raychaudhuri equation:
\be\label{mod-rai}
\f{\ddot a}{a} = - \f{4 \pi G}{3} \, \Sf (\rho_{sc} + 3 P_{sc}) +
\f{1}{2} \f{\dot \Sf}{a} \, \left(\f{\dot a}{\Sf} - \f{\Sf}{\dot
a} \right) ~.
\ee
It is then straightforward to verify,  using the Friedman and the
Raychaudhuri equations, that $\rho$ satisfies the conservation law:
\be\label{mod-cons}
\dot\rho_{sc}+3\frac{\dot a}{a}(\rho_{sc}+P_{sc})=0 ~.
\ee
Defining the modified equation of state as
\be
w_{sc} = P_{sc}/\rho_{sc}
\ee
we therefore have
\be
\rho_{sc}=\frac{\rho_{0sc}}{a^{3(1+w_{sc})}} ~.
\ee

The modified  Klein-Gordon equation can be derived by using
\be
\dot \phi = \{\phi, H_m\} = \f{\partial H_m}{\partial \phi}
\ee
and
\be
\dot p_\phi = \{p_\phi, H_m\} = - \, \f{\partial H_m}{\partial
p_\phi} ~.
\ee
Taking the time derivative of $\dot \phi$ we are led to
\be
\ddot \phi = -3 \, \f{\dot a}{a} \, \left(1 - \f{1}{3} \, \f{d \ln
D_l}{d \ln a} \right) \, \dot \phi - D_l \, \f{\partial
{\cal V}}{\partial \phi} ~.
\ee
For $a  \ll a_\star$, $D_l \ll 1$ and the dynamics of the massive scalar field just
behaves as of the massless scalar field.
It is straightforward to verify  that the energy density
\be
\rho_{sc} = \f{1}{2} \f{\dot \phi^2}{D_l} + {\cal
V}
\ee
and pressure
\be
P_{sc} =
\f{1}{2}\f{\dot \phi^2}{D_l} \, \left(1 - \f{1}{3} \f{d \ln D_l}{d
\ln a} \right) - {\cal V}
\ee
lead to above Klein-Gordon equation through the conservation
equation (\ref{mod-cons}).

The modified equation of state thus becomes
\be
w_{sc} = \f{P_{sc}}{\rho_{sc}} = \f{\dot \phi^2 \left(1 - \f{1}{3}
\f{d \ln
D_l}{d \ln a} \right) - 2 D_l \, {\cal V}}{\dot \phi^2 +
2 D_l \, {\cal V}} ~.
\ee
Using Eq.(\ref{dlta*}) we obtain $w_{sc}$ for $a \ll a_\star$:
\be
w_{sc} \approx 1 - \alpha
\ee
where
\be
\alpha = \f{3 - l}{1 - l} ~.
\ee
The modified equation of state for arbitrary matter can be
similarly found by  following the procedure in
Ref.\cite{ps-disp}. Here one views $\rho_{sc}$ as being obtained
from substituting inverse powers of scale factor as appropriate
powers of $D_l$. If the classical energy density is given by
\be\label{rhoc}
\rho_c= \frac{\rho_0}{a^{3(1+w_c)}}
\ee
then it is easy to see that
\be
\rho_{sc} = \frac{\rho_0}{a^{3(1+w_c)}}D^{w_c}
\ee
Since the latter satisfies a conservation equation, it
evolves according to an expression like (\ref{rhoc}) but with
modified equation of state
\be\label{wscheur}
w_{sc}=w_c\left(1  - \f{1}{3} \f{d \ln D_l}{d \ln a}\right) ~.
\ee
Thus for $a \ll a_\star$ we obtain
\be
w_{sc} \, \approx \, w_c \left(1 - \alpha \right) ~.
\ee
Since $0 < l < 1$, $w_{sc}$ can be easily less than $-1/3$ for
arbitrary matter when $a < a_\star$.

\section{Origin of thermal fluctuations in loop
cosmology}\label{thermal}
In the preceding section we saw that even at the zeroth order
there are modifications to the Friedman dynamics in loop
cosmology. These are sufficient to possibly overcome the no-go
result obtained in classical physics. We start with the simplest
possibility where the only change from the classical physics
appears via (\ref{mod-fried}) and (\ref{mod-rai}). As in Sec. II,
we will be interested in the scales where the modifications coming
from the intrinsic curvature can be ignored in the Friedman
equation. We note that immediately after the bounce, $\rho_{sc}$
becomes dominant over ${\cal S}/a^2$ term and for a proper choice of
initial conditions for matter such a regime can coexist with $a
\ll a_\star$.

Using (\ref{xing}), Eq.(\ref{PT}) modifies to
\be\label{mod-PT}
{\cal P}_\Phi(k) \sim \left[\f{T^2
\rho'_{sc}}{\sqrt{\Sf\rho_{sc}}}\right] ~.
\ee
Since $\rho_{sc} \propto a^{-3(1 + w_{sc})}$ and $\rho \propto
T^\zeta$ we obtain
\be\label{mod-aofT}
a \propto T^{\f{-\zeta}{3(1 + w_{sc})}} ~.
\ee
Using Eq.(\ref{slta*}) we find that for $a \ll a_\star$, ${\cal S} \propto a^3$ and hence
\be\label{mod-nPofT}
\f{d \ln {\cal P}_\Phi}{d \ln T} = \f{(1 + w_{sc})(2 + \zeta) +
\zeta}{2 (1 + w_{sc})} ~.
\ee
Also $k = aH \propto a \sqrt{{\cal S} \rho_{sc}}$ leads to
\be\label{mod-nkofT}
\f{d \ln k}{d \ln T} = \f{\zeta(1 + 3 w_{sc}) - 3 \zeta}{6(1 +
w_{sc})} ~.
\ee
The modifications to the spectral index thus become
\be\label{mod-n-1}
n_S - 1 = \f{d \ln {\cal P}_\Phi}{d \ln k} = 3 \, \f{(1 +
w_{sc})(2 + \zeta) + \zeta}{\zeta(1 + 3 w_{sc}) - 3 \zeta} ~.
\ee
In addition
if we use a semi-classical density for the entropy,
\be
s_{sc}=\frac{S}{V}=\frac{S}{a^3}
\ee
the thermodynamical argument presented in Section~\ref{nogo},
relating $\zeta$ and $w$, is also valid, for the semi-classical
values of these parameters. Thus,
\be
\zeta = 1 + \f{1}{w_{sc}} ~.
\ee
On substituting this  in Eq.(\ref{mod-n-1}), the condition for near scale invariance $n_S \approx 1$
translates to the requirement that $w_{sc} \approx -2/3$ which can
be obtained by an appropriate choice of $l$ parameter and $w_c$ \cite{fn0}.
Therefore we find that at the zeroth order the no-go
conditions for scale invariance of thermal fluctuations can be
overcome by modifications to the gravitational dynamics in loop
cosmology.

However, the calculation presented above is incomplete, because
very little is known about the perturbed Einstein's equations
in loop cosmology, in particular
for $a \ll a_\star$. Preliminary work~\cite{struct} on flat models
and the regime $a \gg a_\star$ suggests that modifications to the
gravitational dynamics influence the growth of fluctuations in a
very non-trivial way. Based on these calculations we classify
below some possible modifications:
\begin{itemize}
\item
In a simplified setting the
Poisson equation inside the horizon could become
\be\label{poisson2}
k^2\Phi=4\pi G a^2 D^I(a)\delta\rho,
\ee
where $I$ is an unknown exponent.
\item It could be that the scale where the fluctuations become dominated
by gravity (and not pressure) is not simply proportional to the horizon
scale $k\sim aH$. For simplicity we shall ignore this possibility:
it relates to varying speed of sound scenarios to be explored elsewhere.
\item It could be that beyond the gravity-driven ``freeze-out'' scale
the potential $\Phi$ continues to evolve and does not freeze-out
as usual. This was proved explicitly in~\cite{struct}
for $a>a_\star$. Here we shall model the evolution of the potential
outside the horizon for $a<a_\star$ as
\be
\Phi\propto a^N
\ee
where $N$ is an exponent to be computed.
\end{itemize}
We stress that these modifications parameterize our ignorance
of the theory {\it but they should be derivable in terms of $j$
and $l$ alone, from first principles}.

Given these novelties
we find that formula (\ref{xing}) gets modified to
\be\label{xing`}
{\cal P}_\Phi(k) \, \sim \, {\left[{a\over k} D^{2I}T^2 \rho_{sc}'\right]}_{k=aH}
\ee
and since $H \propto \sqrt{{\cal S} \rho_{sc}}$ we have at horizon
crossing
\be\label{PT1}
{\cal P}_\Phi(k)\sim {\left[T^2D^{2I} \rho_{sc}'\over \sqrt{\Sf \,
\rho_{sc}}\right]}_{k=aH} ~.
\ee
But because the potential continues to evolve outside the horizon
this is not enough to read off a condition for scale invariance.
Indeed the spectrum left after $a=a_\star$ will be processed into
\be\label{PT2}
{\cal P}_\Phi(k)={\left[T^2D^{2I} \rho_{sc}'\over \sqrt{\Sf \,
\rho_{sc}}a^{2N}\right]}_H
\ee
where all the quantities on the right-hand side are to be computed
when the mode left the horizon, for $k=aH$. The relation
(\ref{aofT}) between temperature and $a$ is unmodified (apart from
replacing $\zeta$ and $w$ by their semi-classical values). Therefore
we can deduce the counterpart of (\ref{nPofT}) as
\be\label{nPofT1}
{d\ln {\cal P}_\Phi\over d\ln T_H}=1+{\zeta\over 2} \left(1 +
\f{1}{3(1+w_{sc})} \left(3 - 12 I \, \alpha + 4 N \right) \right)
\ee
where $T_H$ is the temperature when the mode left the horizon.

Using Eq.(\ref{mod-nkofT}) we then obtain
\be\label{mod-n-1-lqc}
n_S - 1 = \f{3(1 + w_{sc})(\zeta + 2) + \zeta(3 - 12 I \alpha + 4
N)}{\zeta(1 + 3 w_{sc}) - 3 \zeta} ~.
\ee
The condition for near scale invariance ($n_S \approx 1$) then implies 
$3(1 + w_{sc})(\zeta + 2)  \approx - \zeta(3 - 12 I \alpha + 4
N)$ leading to 
\be \label{mod-w-cond}
w_{sc} \approx -\f{2}{3} - \f{4}{9} \left(N - 3 I \alpha \right)
~.
\ee
This equation can be viewed as a first order improvement over our
zeroth order calculation which led to $w_{sc} \approx -2/3$.

\section{Conclusions}\label{concs}
Loop quantum cosmology has the potential to relate observational
physics and quantum gravity, allowing concrete calculations to be
made in the quantum gravity regime as long as a minisuperspace
approximation is assumed to be valid. The approach is known to
modify the equation of state of ordinary matter, thereby
permitting a solution of the horizon problem without resorting to
scalar fields. It is then natural to ask whether in such scenarios
thermal fluctuations could be behind the observed structure of the
Universe. In order to analyze this issue we have assumed in this
paper that physics learned in the mini-superspace approximation (in
the sense of modifications to inverse volume terms) will not
change qualitatively. There are positive indications for this hope
from ongoing work~\cite{ham-pert}, but we stress this important
caveat in our analysis.

We showed that prima facie we are confronted by a no-go result in
classical physics, pointing to $n_S=4$ in {\it all} such
scenarios. This can be derived assuming only the Einstein
equations and a basic thermodynamics relation. The fact that the
zeroth order, background, Einstein equations are also modified in
loop cosmology allows us to bypass this negative result, pointing
to the region of parameter space where  $n_S\approx 1$ is
realized. This occurs for the semi-classical equation of state
$w_{sc}\approx -2/3$. However, before this requirement can be
converted into a constraint upon the free parameters of the theory
($j$ and $l$), a number of important details have to be worked
out. We close with an executive summary of what is still missing
in the theory:
\begin{itemize}
\item A solid quantization in the regime of large $j$, necessary for
a full understanding of an extended period with $a<a_\star$. This has
to be accomplished for closed models as for non-compact flat models
the physical meaning of $a_\star$ makes little sense.
\item A study of the perturbed Einstein equations along the lines
of that carried out in~\cite{struct},
but valid for $a<a_\star$.
\item A concrete prediction for the spectrum
of gravitational waves (tensor modes) completely ignored in this paper.
\end{itemize}
We believe that once this task list is completed we shall be able
to place solid observational constraints upon loop quantum cosmology.

We conclude with a final remark on the role of higher $j$ terms in
both the gravitational and the matter parts of the Hamiltonian. In
various LQC phenomenology papers one has often ignored the
modification to the gravitational part (constituted by ${\cal
S}$). Such an ad-hoc analysis is similar to taking different
metrics in the gravity and matter parts of the Einstein equations
in GR. As a purely academic exercise we can perform such an
analysis and it turns out that the zeroth order calculation in
Section \ref{thermal} does not go through; instead one reproduces
the no-go result for scale invariance of classical
physics~\cite{fn3}. But by
consistently incorporating modifications arising for high values
of $j$ in both the gravitational and matter parts of the
constraint, the no-go obstacle is removed even at the zeroth level
of calculation. This is an important lesson for loop cosmology
phenomenology, showing the non-trivial features of high $j$. We
believe this opens an interesting avenue for re-examining various
interesting ideas (e.g., Refs.\cite{lqc-old,struct}).

\section*{Acknowledgments}

We thank Stephon Alexander for useful discussions in early period of
this work and Lee Smolin for helpful comments. Research at PI is
supported in part by the Government of Canada through NSERC and by
the Province of Ontario through MEDT. PS is supported by the NSF
grant PHY-0456913 and the Eberly research funds of Penn State.

\begin{appendix}

\section{Measures of structure}\label{measures}

For
discrete Fourier modes we define, for any quantity $g$, the
dimensionless (or curly) power spectrum ${\cal P}_g(k)$ as:
\be
{\cal P}_g(k)={V\over (2\pi)^3}4\pi k^3{\langle
|g_\vk|^2\rangle} \ee %
or occasionally the non-curly one as:
\be
P_g(k)=V {\langle |g_\vk|^2\rangle}={2\pi^2\over k^3} {\cal
P}_g(k) ~.\ee
The latter is often only used for $\delta$ and using the Poisson equation 
we have the following alternative definition of the the spectral
index
\be
P_\delta(k)=A^2k^n ~.
\ee

The Fourier transform can be introduced noting that $\Delta
k=2\pi/L$ so that: %
\be \int d^3 k\approx {(2\pi)^3\over V}\sum_{\mathbf k} ~.\ee
Then with:
\bea g(\vx)&=&{1\over (2\pi)^{3/2}}\int d\vk \, g(\vk)e^{i\vk\cdot \vx}\\
 \delta (\vk)&=&{1\over (2\pi)^3}\int d\vx \, e^{i\vk\cdot \vx}\\
g(\vk)&=&{1\over (2\pi)^{3/2}}\int d\vx \, g(\vx) e^{-i\vk\cdot
\vx}
  \eea
we have %
\bea g_\vk&\approx&g(\vk){(2\pi)^{3/2}\over V}\\
\delta_{\vk\vk '}&\approx&{(2\pi)^3\over V}\delta(\vk-\vk ')\eea
and so we find the alternative {\it and equivalent} definition for
the power spectrum:
\be%
{\langle g(\vk)g^\star(\vk ')\rangle}={2\pi^2\over k^3}{\cal
P}_g(k)\delta(\vk -\vk ')~.\ee
The position space variance, with either definition, can be
written: %
\be \sigma^2_g={\langle g^2(\vx)\rangle}=\int {dk\over k}{\cal
P}_g(k)~.\ee
The filtered position-space variance is also used. It's based on
the smoothed field \bea g(R,x)&=&{1\over V_R}\int
g(x')W(|x-x'|/R)d^3x'\\
 V_R = &&\hskip-0.5cm \int d^3x W(x/R)=4\pi R^3\int
y^2W(y) dy
\eea where $R$ is the smoothing scale and $W$ can be, say, a
Gaussian or a top hat. (Do not confuse $V_R$ with the large $V$
used in the discrete Fourier series.) Then the ``sigma-squared''
on scale $R$ is
\be \sigma^2_g(R)={\langle g^2(R,x)\rangle}=\int
{dk\over k}W^2(kR){\cal P}_g(k) \ee
where we used the convolution theorem and $W(kR)$ is the Fourier
transform of $(2\pi)^{3/2}W(\vx/R)/V_R$, that is, it's normalized
so that $W(kR)=1$ at $k=0$, then falling off at $k_R\sim 1/R$. We
can then write
approximately %
\be\label{sigPapprox} \sigma^2_g(R)\approx {\cal
P}_g(k_R) \ee
for $k_R=1/L$, since that's where the integrand peaks.

\section{Some Basics of Loop Quantum Cosmology}\label{lqc-app}

This appendix aims to summarize some key features of loop quantization
of cosmological models (for details see for example Refs.\cite{aps2,apsv}). We first
demonstrate the way the classical constraint is cast in terms of
elementary loop variables -- holonomies of connection $h_e(A) =
{\cal P} \exp(\int_e A)$ and triads $E^a_i$. We also show the way 
$j$ and $l$ parameters  
and the resulting
modifications in the improved quantization of LQC \cite{aps2} to classical
GR originate.

For simplicity we start with quantization of flat isotropic and
homogeneous FRW spacetime. In this setting the underlying
symmetries lead to simplified connection $c$ and triad $p$\,:
\be A_a^i = c\,V_o^{-\f{1}{3}} {}^{o\!}\omega_a^i, \quad {\rm and} \quad E^a_i
= p\, V_o^{-\f{2}{3}} \sqrt{q_o}\,\, {}^{o\!}e^a_i \ee
where $({}^{o\!}\omega_a^i, {}^{o\!}e^a_i)$ are a set of
orthonormal co-triads and triads compatible with the flat fiducial
metric ${}^{o\!}q_{ab}$. $V_o$ is the volume of the cell ($\tt V$) used to
define a symplectic structure with respect to ${}^{o\!}q_{ab}$.
The variables $c$ and $p$ are canonical conjugate satisfying
Eq.(\ref{cpbp}).

The gravitational and matter  parts of the constraint are given by
\be
C_{\mathrm{grav}} = - \g^{-2} \int_{\tt V} \, d^3 x \epsilon_{ijk}
\, e^{-1} E^{ai} E^{bj} F^k_{ab} ~
\ee
and
\be
C_m = 8 \pi G \, \f{p_\phi^2}{|p|^{3/2}} 
\ee
where for simplicity we consider a massless scalar field.
The modulus sign arises because of two possible orientations of the triad, the choice of which has no physical consequences unless we choose spinor fields.
 $F^i_{ab}$ denotes the curvature of connection and
\be\label{dete}
e = \sqrt{|\det E|} = \left(\f{1}{6} |\epsilon^{lmn}
\epsilon_{ijk} \, E^a_l E^b_m E^c_n|\right)^{1/2} ~.
\ee
To write the constraint in terms of holonomies and triads, the following
identities of classical phase space are very useful \cite{qsd5}:
\be\label{firstiden}
\f{1}{8 \pi G \g} \, \{A_l^d,\epsilon^{ijk} \epsilon_{abc} \,
E^a_i E^b_j E^c_k\} = 3 \, \epsilon^{ijl} \epsilon_{abd}  \, E^a_i
E^b_j~,
\ee
\be\label{seciden}
\f{\{A^i_a,V\}}{V^n} = \f{\{A^i_a,V^{(1-n)}\}}{1 - n}
\ee
and
\be\label{thirdiden}
e^i_a = \f{1}{4 \pi G\g} \{A^i_a, V\} ~.
\ee
Eq.(\ref{firstiden}) leads to
\bq
\epsilon_{ijk} \, e^{-1} E^{ai} E^{bj} F^k_{ab} &=& \nonumber
\f{1}{8 \pi G \g} \, e^{-1} \, \epsilon^{abc} \, \{A_c^i, V^2 \}
\, F_{abi} \\
&=& \f{1}{4 \pi G \g} \, \epsilon^{abc} \, \{A_c^i, V \} \,
F_{abi} ~.
\eq
Here $V = |p|^{3/2} = a^3$ denotes volume of the cell with respect to the physical metric
$V = V_o \sqrt{|\det E|}$ (for simplicity we put $V_o = 1$ from
now on).

We then express the connection in terms of the holonomy by tracing over the
holonomies in a $j$ representation. For $j=1/2$, using
\be\label{trace}
\mathrm{Tr}(\tau_i \tau^j) = -\f{1}{3} \, j (j + 1) (2j+1) \,
\delta_i^j ~,
\ee
it is straightforward to obtain
\be\label{doubE}
\epsilon_{ijk} \, e^{-1} E^{ai} E^{bj} = \sum_k \,
\f{\epsilon^{abc} \, {}^{o\!}\omega_c^k}{2 \pi G \g \lambda} \,
\mathrm{Tr}(\hkl\{\hkli,V\}\tau_i) ~.
\ee
Here $h_k^{(\lambda)}$ is the holonomy of the connection $c$ along
the edge $\lambda {}^{o\!}e^a_k$
\be
h_k^{(\lambda)} = \cos \f{\lambda}{2} \,{\mathbb{I}} \,+ 2 \, \sin
\f{\lambda c}{2} \, \tau_k
\ee
where ${\mathbb I}$ is the identity matrix, $\tau_k = - i
\sigma_k/2$ and $\sigma_k$ are the
Pauli spin matrices.

The curvature components can be obtained by considering holonomies
around a closed square loop $\Box_{ij}$:
\be \label{F} F_{ab}^k\, = \, -2\,\lim_{Ar_\Box
  \rightarrow 0} \,\, \mathrm{Tr}\,
\Bigg[\left(\f{h^{(\lambda)}_{\Box_{ij}}-1 }{\lambda^2} \right)
\,\, \tau^k \Bigg]\, {}^{o\!}\omega^i_a\,\, {}^{o\!}\omega^j_b\, .
\ee
Due to the inherent quantum nature of geometry in loop quantization the area
$\lambda^2 |p|$ is shrunk to the minimum eigenvalue of the area
operator $\Delta = (2\sqrt{3} \pi \g) \lp^2$ in LQG \cite{aps2}.
This leads to a constraint $\lambda^2 |p| = \Delta$.
$C_{\mathrm{grav}}$ can then be obtained by combining
(\ref{doubE}) and (\ref{F}).

The matter part contains inverse powers of $\det E$. To write them in
 terms of holonomies we use the identity $|\det e^i_a| = |\det E|^{1/2}$
and Eq.(\ref{thirdiden})
\bq\label{invsqdete}
\f{1}{\sqrt{|\det E|}} &=& \nonumber \f{\det e^i_a}{(\sqrt{|\det
E|})^2} = \f{1}{6} \, \f{\epsilon^{abc} \epsilon_{ijk} e^i_a e^j_b
e^k_c}{\det E} \\
&& \nonumber \hskip-1.5cm = \f{(4 \pi G \g)^3}{6} \,
\epsilon^{abc} \epsilon_{ijk} \, \{A_a^i,V^{1/3}\}
\{A_b^j,V^{1/3}\}\{A_c^k,V^{1/3}\} ~. \\
&&
\eq
Expressing connection in terms of holonomies, in $j=1/2$
representation inverse scale factor becomes
\be\label{inv-a}
\f{{\mathrm{sgn}(p)}}{|p|^{1/2}} = \bigg[\f{4}{8 \pi
G \gamma \lambda } {\mathrm{Tr}}\, \sum_k \, \tau^{k} \,\hkl
\{\hkli,\,
V^{1/3}\} \bigg] ~.
\ee
Note that in obtaining (\ref{invsqdete}) we could have multiplied
by $(\det e^i_a/\sqrt{\det E})^m$ where $m$ has a value such that we
obtain positive powers of the volume in the final expression of
$1/{\sqrt{|\det E|}}$. This happens when $m > 1/2$. We are thus
led to the $l$ ambiguity in the expressions for inverse triads in
the matter part of the constraint and Eq.(\ref{inv-a}) generalizes
to
\be
\f{{\mathrm{sgn}(p)}}{|p|^{1/2}} = \bigg[\f{2}{8 \pi G \gamma l
\lambda } {\mathrm{Tr}}\, \sum_k \, \tau^{k} \,\hkl \{\hkli,\,
V^{2l/3}\} \bigg]^{1/2(1 - l)} ~
\ee
where $l = 1 - 1/(2m)$ and $0 < l < 1$.

With $C_{\mathrm{grav}}$ and $C_m$ written in form of holonomies
and triads, we quantize the theory and are led to a non-singular difference
equation with uniform discretization in eigenvalues of volume
operator \cite{aps2}:
\be
\hat V |v\rangle = \tilde \beta^3 \, |v| |v\rangle
\ee
with
\be\label{alpha}
\tilde\beta = \left(\f{8 \pi \lp^2 \g}{6}\right)^{1/2}
K^{-1/3},  \, ~~ \, K = \f{2\sqrt{2}}{3\sqrt{3\sqrt{3}}} ~.
\ee
Unlike the old quantization in LQC (where the difference equation was of uniform discretization in eigenvalues
of triad \cite{abl,aps1}), the evolution has the correct classical
limit for arbitrary matter content and quantum gravitational effects set
in when curvature becomes of the order Planck. 
 Study of backward evolution of semi-classical states
peaked at late times on trajectories of a large classical universe
shows a generic bounce when $\rho =
\rho_{\mathrm{crit}} = 0.82 \rho_{\mathrm{Pl}}$ \cite{aps2}.
In this quantization (for $j=1/2$)
modifications originating from $F_{ab}^i$ terms dominate over
those containing $1/\sqrt{\det E}$ in both gravitational and
matter parts of the constraint. Investigations of closed models
yield similar results \cite{apsv}.

We now provide the expressions of the eigenvalues of operators
corresponding to $e^{-1} E^{ai} E^{bi}$ and $1/{\sqrt{\det E}}$
for higher $j$, which can be derived in analogy with (\ref{doubE})
and (\ref{inv-a}), using Eq.(\ref{trace}). We have:
\be \label{sj}
s_j = - \, \f{9 ~ \tilde\beta^3 ~ K^{2/3}}{8 \pi \lp^2 \g
j(j+1)(2j+1)} \, |v|^{1/3} \, \sum_{r = -j}^{j} \, r |v - 2 r|
\ee
and
\be \label{djl}
d_{j,l} = \Bigg[\f{27 ~ \tilde\beta^{2l} ~ K^{2/3}}{16 \pi \lp^2 \g
l j(j+1)(2 j + 1)} |v|^{1/3} \, \sum_{r = -j}^j \, r |v +
2r|^{2l/3} \Bigg]^{\f{3}{2(1-l)}}.
\ee
For higher $j$, Eqs.(\ref{sj}) and (\ref{djl})
can be approximated by 
\be\label{sjdj}
s_j = \Sf(q) \, a, ~~~ {\mathrm{and}} ~~~ d_{j,l}(q) = D_l(q) a^3
\ee
with
\be\label{astar}
q := (a/a_\star)^3 , \, \, a_\star = (2j)^{1/3} \, \tilde\beta ~,
\ee
\bq\label{slta*}
\Sf(q) = && \hskip-0.3cm\nonumber \f{1}{4} \Bigg[2\left((q+1)^3 -
|q-1|^3\right)\\ &-& 3 q \left((q+1)^2 - \mathrm{sgn}(q-1)
|q-1|^2\right) \Bigg]
\eq
and
\bq\label{dlta*}
D_l(q) &=& \nonumber \Bigg[\f{27 ~ |q|^{1 - \f{2l}{3}}}{8 l}\,
\Big\{\f{1}{l + 3} \left((q + 1)^{\f{2(l+3)}{3}} -
|q-1|^{\f{2(l+3)}{3}} \right)  \\
 &&\hskip-1cm - \f{2q}{2l + 3} \left((q+1)^{\f{2(l+3)}{3}} -
\mathrm{sgn}(q-1) |q-1|^{\f{2(l+3)}{3}} \right)\Big\}
\Bigg]^{\f{3}{2(1-l)}} 
\eq
following the analysis for old
quantization \cite{bojo-amb,kv-ham}.
In the regime when $a \ll a_\star$
\be
\Sf(q) \approx \f{3}{2} \, \left(\f{a}{a_\star}\right)^3
\ee
and
\be
D_l(q) \approx \left(\f{9}{2 l + 3}\right)^{3/(2(1-\,l))} \,
\left(\f{a}{a_\star}\right)^{3(3 - \,l)/(1 - \,l)} ~.
\ee
For $a \gg a_\star$, $\Sf(q) \approx 1$ and $D_l(q) \approx 1$.

Extensive numerical simulations of backward evolution of semi-classical states at late times have shown that
in LQC an effective
Hamiltonian which provides an excellent approximation to the
underlying quantum dynamics can be written. For $j=1/2$, the
effective Hamiltonian for flat \cite{aps2} and closed model
\cite{apsv} is
\be\label{heff0} \heff =  \f{C_{\rm eff}}{16\pi G} = -
\f{3 s_j \,\sin^2(\lambda c)}{8 \pi G \gamma^2 \lambda^2}
  +  d_{j,l} \f{p_\phi^2}{2} \, ~
\ee
and
\be\label{heff1} \heff = - \f{3 s_j \, \left(\sin(\lambda c)\sin(\lambda (c - 1)) + (1 + \g^2)\right)}{8 \pi G \gamma^2 \lambda^2}
   +  d_{j,l} \f{p_\phi^2}{2} \, ~
\ee
respectively. Here $\lambda = \lambda(p) = (\Delta/|p|)^{1/2}$.
The $\sin(\lambda c)$ terms arise from field strength part and
are responsible for  $\rho^2$ modifications of the
Friedman equation \cite{aps2,apsv}.

For higher $j$ a
quantization procedure has been proposed which indicates
resolution of singularity \cite{kv-ham}; however a complete
quantization is still lacking. Considering higher values of $j$ in
non-compact flat models has the problem of relating the scale at
which modifications to $1/\sqrt{\det E}$ terms become important to
any physical scale (The scale at which these modifications become
important depends on the choice of fiducial cell, for details see
Refs.\cite{kv-k-1,aps2}).
In the closed models this scale is provided by the intrinsic
curvature.
If we consider higher $j$ in closed models then, as
for $j=1/2$, modifications arise both from $F_{ab}^i$ and
$1/\sqrt{\det E}$ terms. However, in this case we can have a
regime in which modifications coming from the latter dominate the
former and still lead to a non-singular bounce. When $\sin (\lambda c)
\rightarrow \Delta^{1/2} c/a$ and $\sin(\lambda (c-1)) \rightarrow
\Delta^{1/2} (c-1)/a$, the modifications coming from $F_{ab}^i$
can be considered small.
In this regime $H = \dot a/a \ll \Delta^{-1/2} \sim
M_{\mathrm{Pl}}$ and energy density is small.
An effective Hamiltonian (\ref{effhamk1}) can
then be obtained by following the procedure outlined for higher
$j$ \cite{kv-ham}.

\end{appendix}


\begin{thebibliography}{99}
\bibitem{fn2} An alternative way to obtain effective energy density in
loop cosmology is by construction of a local energy density operator
\cite{ps-disp}. This definition does not lead to energy when
integrated over volume and creates additional subtleties
including possible violation of extensiveness of energy.
\bibitem{fn0} Trivial solution of Eqs.(\ref{mod-n-1}) and 
(\ref{mod-n-1-lqc}) is obtained for $\zeta = 0$ for which there are no
thermal fluctuations.
\bibitem{fn3} The no-go, however,  can be avoided by
considering modifications to the perturbed equations.
\end{thebibliography}
\end{document}